\documentclass[12pt]{article}
\usepackage{jhep-mod}
\usepackage{bm}
\usepackage{amssymb,amsmath,amsthm}
\usepackage{mathrsfs}
\usepackage[utf8]{inputenc}
\usepackage{enumerate}
\hypersetup{colorlinks=true} 
\usepackage{xcolor}
\usepackage{appendix}
\usepackage{graphicx}
\usepackage{dcolumn}
\usepackage{bm}
\usepackage{multirow}
\usepackage{float}
\usepackage{tikz}
\usepackage[normalem]{ulem}
\definecolor{purple}{rgb}{1,0,1}
\definecolor{lime}{HTML}{A6CE39} % needs xcolor

\newcommand{\red}[1]{{\slshape\color{red} #1}}
\newcommand{\blue}[1]{{\slshape\color{blue} #1}}
\newcommand{\purple}[1]{{\slshape\color{purple} #1}}
\newcommand{\orcidicon}{%
	\begin{tikzpicture}
	\draw[lime, fill=lime] (0,0) 
		circle [radius=0.16] 
		node[white] {{\fontfamily{qag}\selectfont \tiny ID}};
	\draw[white, fill=white] (-0.0625,0.095) 
		circle [radius=0.007];
	\end{tikzpicture}
	\hspace{-2mm}
}
\newcommand\orcidMatt{{\href{https://orcid.org/0000-0003-1088-6485}{\orcidicon}}}
\begin{document}
%========================================================

\title{\huge The Kiselev black hole is neither perfect fluid, nor is it quintessence}

\author{
\Large Matt Visser\orcidMatt}
%========================================================
%========================================================
%========================================================
%========================================================
\affiliation{
School of Mathematics and Statistics, Victoria University of Wellington, \\
\null\qquad PO Box 600, Wellington 6140, New Zealand}
%========================================================
%========================================================
\emailAdd{matt.visser@sms.vuw.ac.nz}
%========================================================
%========================================================

\abstract{
\parindent0pt
\parskip7pt
The Kiselev black hole spacetime,
\begin{equation*}
ds^2 = - \left(1-{2m\over r} - {K\over r^{1+3w}} \right) dt^2 + {dr^2\over1-{2m\over r} - {K\over r^{1+3w}}} 
+ r^2 \,d\Omega_2^2,
\end{equation*}
is an extremely popular toy model, with over 200 direct and indirect citations as of 2019. Unfortunately, despite repeated assertions to the contrary, this is not a perfect fluid spacetime. The relative pressure anisotropy and average pressure are  easily calculated to satisfy
\begin{equation*}
\Delta = {\Delta p\over \bar p} = {p_r - p_t \over {1\over3} (p_r+2p_t)} =- {3(1+w)\over 2 w};
\qquad\qquad
{\bar p\over \rho} = {{1\over3} (p_r + 2p_t)\over \rho} = w.
\end{equation*}
The relative pressure anisotropy $\Delta$ is generally a non-zero constant, (unless $w=-1$, corresponding to Schwarzschild-(anti)-de~Sitter spacetime). Kiselev's original paper was very careful to point this out in the calculation, but then in the discussion made a somewhat unfortunate choice of terminology which has (with very limited exceptions) been copied into the subsequent literature. 
Perhaps worse, Kiselev's use of the word ``quintessence'' does not match the standard usage in the cosmology community, leading to another level of unfortunate and unnecessary confusion. Very few of the subsequent follow-up papers get these points right, so a brief explicit comment is warranted.

\bigskip
{\sc Date:} 29 August 2019; \LaTeX-ed \today

%\bigskip
%{\sc ArXiv:} 1908.nnnnn


\bigskip
{\sc PhySH:} \\
Gravitation; Classical black holes; Fluids \& classical fields in curved spacetime.
}

%\notoc
\maketitle

%-----------------------------------------------------------------------------------------------------
\definecolor{purple}{rgb}{1,0,1}
\renewcommand{\red}[1]{{\slshape\color{red} #1}}
\renewcommand{\blue}[1]{{\slshape\color{blue} #1}}
\renewcommand{\purple}[1]{{\slshape\color{purple} #1}}
%-----------------------------------------------------------------------------------------------------
\newcommand{\redm}[1]{{\color{red} #1}}
\newcommand{\bluem}[1]{{\color{blue} #1}}
\newcommand{\purplem}[1]{{\color{purple} #1}}
%-----------------------------------------------------------------------------------------------------

%========================================================
\def\tr{{\mathrm{tr}}}
\def\diag{{\mathrm{diag}}}
\parindent0pt
\parskip7pt

\def\rocco{r_\mathrm{\scriptscriptstyle OCCO}}
\def\ricco{r_\mathrm{\scriptscriptstyle ICCO}}
\def\rosco{r_\mathrm{\scriptscriptstyle OSCO}}
\def\risco{r_\mathrm{\scriptscriptstyle ISCO}}
%========================================================
\section{Introduction}
%========================================================
%---------------------------------------------------------------------------------------------------------------------------------------------
\label{S:intro}
%---------------------------------------------------------------------------------------------------------------------------------------------

Kiselev's black hole spacetime~\cite{Kiselev:2002},
\begin{equation}
ds^2 = - \left(1-{2m\over r} - {K\over r^{1+3w}} \right) dt^2 + {dr^2\over1-{2m\over r} - {K\over r^{1+3w}}} 
+ r^2 \,d\Omega_2^2,
\end{equation}
is a remarkably popular toy model. Directly and indirectly, Kiselev's model has accum\-ulated over 200 citations, with over 150 of the citing articles being published.
One reason for this model's popularity is its generality: $w=0$ corresponds to Schwarzschild, $w=1/3$ corresponds to Reissner--Nordstr\"om, and $w=-1$ corresponds to Schwarzschild-(anti)-de~Sitter (Kottler). 
Unfortunately a very large fraction of the subsequent follow-up papers discussing Kislev's model get basic aspects of the physics wrong. Despite (very) many assertions to the contrary, the Kiselev spacetime is not a perfect fluid spacetime, neither does it have anything to do with the cosmologist's notion of quintessence. 

Perhaps the fastest way to see something is wrong with the \emph{terminology} (without having to do a calculation) is to consider the special case $w=1/3$ with $K=-Q^2$ (that is, Reissner--Nordstr\"om), and note that the electromagnetic field is not a perfect fluid, nor can the electromagnetic field meaningfully be described as quintessence. 

\clearpage
Despite these \emph{terminological} issues, the Kiselev black hole does have some interesting physical and mathematical properties, and does merit investigation --- as long as one does so carefully, and uses terminology in a manner consistent with the broader astrophysical and general relativity communities. 

%========================================================
\section{Stress-energy}
%========================================================

Working in an orthonormal frame it is easy to see
\begin{equation}
G_{\hat t\hat t} = - G_{\hat r \hat r} = -{3Kw\over r^{3(1+w)}};
\qquad\qquad
G_{\hat\theta\hat\theta} = G_{\hat\phi \hat\phi} = - {3Kw(1+3w)\over2r^{3(1+w)}}.
\end{equation}
Therefore 
\begin{equation}
\rho = - p_r = -{3Kw\over8\pi r^{3(1+w)}};
\qquad\qquad
p_t = - {3Kw(1+3w)\over16\pi r^{3(1+w)}}.
\end{equation}
This is not isotropic, so it is not a perfect fluid. 
For the average pressure we have
\begin{equation}
\bar p = {p_r +2p_t\over3} =-{3Kw^2\over8\pi r^{3(1+w)}};
\qquad\qquad
{\bar p\over\rho} = w.
\end{equation}
While such an average pressure can always be defined, doing so does not magically convert an anisotropic stress-energy into a perfect fluid.
Indeed for the pressure ratio and relative pressure anisotropy we explicitly have
\begin{equation}
{p_t\over p_r} = - {1+3w\over2}; 
\qquad\qquad 
\Delta ={\Delta p\over\bar p} = {p_r-p_t\over\bar p} = -{3(1+w)\over2w}. 
\end{equation}
Note that this basic Kiselev spacetime has the interesting feature that both  of the ratios $p_t/p_r$ and $\Delta$ are position-independent constants.
However, since for $w\neq-1$ we have both $p_t/p_r \neq 1$ and $\Delta\neq0$,  this is certainly not a perfect fluid spacetime. 

Unfortunately, mistakenly mis-identifying anisotropic stress-energies as perfect fluids has a distressingly long history in general relativity~\cite{Delgaty:1998}. (This was unfortunate but  perhaps understandable in the days before computer-based symbolic algebra packages, when all curvature calculations had to be done by hand~\cite{Delgaty:1998}, it is considerably less understandable in the present day.) In the present context, very few of the follow-up papers to Kiselev's original result~\cite{Kiselev:2002} have been careful in this regard --- for a notable exception see reference~\cite{Cvetic:2016} where the authors very carefully and explicitly specify the stress-energy tensor being used, and pointedly do not refer to this spacetime as a perfect fluid spacetime. 

%========================================================
%========================================================

Note that because the Kiselev spacetime is static and spherically symmetric it \emph{will} be possible to model the matter distribution by some linear combination of perfect fluid plus scalar field (with spacelike gradient) and electromagnetic field~\cite{Boonserm}, but that is a very different statement from the assertion that it is a perfect fluid spacetime. 

Let us turn now to the word ``quintessence'' as used within the cosmology community. At its most basic ``quintessence'' refers to a scalar field with a timelike gradient, see for instance~\cite{Caldwell:1997,Zlatev:1998,Sahni:2002,Padmanabhan:2002,Sahni:2004,Copeland:2006}.
In particular, the stress-energy tensor associated with quintessence is that of a zero-vorticity perfect fluid. Therefore the Kiselev spacetime does not represent quintessence in the sense that this word is normally used within the cosmology community. 
Even those cosmological models that seek to break quintessence away from the scalar field framework~\cite{Capozziello:2003}, still retain a perfect fluid stress-energy tensor, and so are intrinsically incompatible with the matter distribution in the Kiselev spacetime. 

Now on the one hand this is just a matter of \emph{terminology}, on the other hand terminology matters --- only if there is widespread agreement on the meaning of the words being used can useful scientific communication take place.

%---------------------------------------------------------------------------------------------------------------------------------------------
\section{Generalized Kiselev black holes I}
%---------------------------------------------------------------------------------------------------------------------------------------------
Consider now a slightly generalized two-component version of Kiselev spacetime~\cite{Kiselev:2002}
\begin{equation}
ds^2 = - \left(1-{2m\over r} - {K_1\over r^{1+3w_1}} -{K_2\over r^{1+3w_2}} \right) dt^2 
+ {dr^2\over1-{2m\over r} - {K_1\over r^{1+3w_1}} -{K_2\over r^{1+3w_2}} } 
+ r^2 \,d\Omega_2^2.
\end{equation}
This two-component generalization is already enough to see interesting new effects. 

For the stress-energy we now have
\begin{equation}
\rho = - p_r = -{3(K_1\,w_1\, r^{-3w_1} +K_2\,w_2\, r^{-3w_2}) \over8\pi r^3},
\end{equation}
and
\begin{equation}
p_t = - {3(K_1\,w_1\,(1+3w_1)\, r^{-3w_1} + K_2\,w_2\,(1+3w_2)\, r^{-3w_2}     )\over16\pi r^3}.
\end{equation}
For the average pressure we now have
\begin{equation}
\bar p = {p_r +2p_t\over3} =-{3(K_1\,w_1^2\,r^{-3w_1} + K_2\,w_2^2\,r^{-3w_2}) \over8\pi r^3},
\end{equation}
while we now define
\begin{equation}
w_\mathrm{effective} := {\bar p\over\rho} =  
{K_1\,w_1^2\,r^{-3w_1} + K_2\,w_2^2\,r^{-3w_2}\over K_1\,w_1\, r^{-3w_1} +K_2\,w_2\, r^{-3w_2}}. 
\end{equation}
Note that $w_\mathrm{effective}$ is no longer position independent; it can however be viewed as a position-dependent weighted average of $w_1$ and $w_2$.

For the relative pressure anisotropy we now have
\begin{equation}
\Delta ={\Delta p\over\bar p} = {p_r-p_t\over\bar p} = {p_r - (3\bar p - p_r)/2\over\bar p} =  {3\over2} \, {p_r - \bar p\over\bar p}
= -{3\over2} \,  {\rho + \bar p\over\bar p},
\end{equation}
whence
\begin{equation}
\Delta  = -{3(1+w_\mathrm{effective})\over2w_\mathrm{effective}}. 
\end{equation}
Note the the relative pressure anisotropy is also no longer position independent. 
If one wishes to be explicit
\begin{equation}
\Delta  = -{3\over2} 
\left(1+   
{K_1\,w_1\,r^{-3w_1} + K_2\,w_2\,r^{-3w_2}\over K_1\,w_1^2\, r^{-3w_1} +K_2\,w_2^2\, r^{-3w_2}}
\right).
\end{equation}
So while one can still do quite simple calculations in this two-component model, one has lost one of the most compelling features of the simple one-component model --- the relative pressure anisotropy is now a somewhat complicated function of position.

%---------------------------------------------------------------------------------------------------------------------------------------------
\section{Generalized Kiselev black holes II}
%---------------------------------------------------------------------------------------------------------------------------------------------
Now consider the $N$-component generalized Kiselev spacetime~\cite{Kiselev:2002}
\begin{equation}
ds^2 = - \left(1 - {\sum_{i=1}^N K_i \,r^{-3w_i}\over r} \right) dt^2 
+ {dr^2\over 1 - {\sum_{i=1}^N K_i \,r^{-3w_i}\over r}}
+ r^2 \,d\Omega_2^2.
\end{equation}
Any Schwarzschild mass term that might be present has now been absorbed into one of the $K_i$ by setting the corresponding $w_i$ to zero.
Effectively one is defining a position-dependent mass function $m(r)$ by setting
\begin{equation}
\label{E:mass}
2\, m(r) = \sum_{i=1}^N K_i \,r^{-3w_i},
\end{equation}
and considering a metric of the form~\cite{Jacobson:2007}
\begin{equation}
\label{E:normal}
ds^2 = - \left(1 - {2m(r)\over r} \right) dt^2 
+ {dr^2\over 1 - {2m(r)\over r}}
+ r^2 \,d\Omega_2^2.
\end{equation}
Spacetime metrics of this form have very special properties~\cite{Jacobson:2007}, and
it is then an utterly standard calculation to show
\begin{equation}
\rho = - p_r = {m'(r)\over4\pi r^2},
\qquad
\hbox{and}
\qquad
p_t = -{m''(r)\over8\pi r}.
\end{equation}
For the average pressure we now have
\begin{equation}
\bar p = {p_r +2p_t\over3} =-{m'(r) + r m''(r)\over12\pi r^2};
\qquad\qquad
w_\mathrm{effective}:= {\bar p\over\rho} = -{1\over3} - {r m''(r)\over 3 m'(r)}.
\end{equation}
For the ratio of pressures we now have
\begin{equation}
{p_t\over p_r} =  {r\,m''(r)\over2m'(r)} = - {3 w_\mathrm{effective}+1\over2},
\end{equation}
and so for the relative pressure anisotropy 
\begin{equation}
\Delta ={\Delta p\over\bar p} = {p_r-p_t\over w_\mathrm{effective}\rho} = - {1-(p_t/p_r)\over w_\mathrm{effective}} 
= -{3(1+w_\mathrm{effective})\over2w_\mathrm{effective}}. 
\end{equation}
In general $w_\mathrm{effective}$, the ratio of pressures $p_t/p_r$, and the relative pressure anisotropy $\Delta$ are now all position dependent. Note that these key properties follow directly from the general form of the metric as given in (\ref{E:normal}) and do not need the explicit form of the mass function $m(r)$ as given in (\ref{E:mass}). 

However, if one wishes to be explicit and keep all the individual $K_i$ and $w_i$ visible, then it is easy to see that for the stress-energy
\begin{equation}
\rho = - p_r = -{3 \sum_{i=1}^N K_i\,w_i\, r^{-3w_i} \over8\pi r^3},
\end{equation}
and
\begin{equation}
p_t = - {3  \sum_{i=1}^N K_i\,w_i\,(1+3w_i)\, r^{-3w_i}  \over16\pi r^3}.
\end{equation}
For the average pressure we now have
\begin{equation}
\bar p = {p_r +2p_t\over3} =-{3  \sum_{i=1}^N K_i\,w_i^2\,r^{-3w_i} \over8\pi r^3},
\end{equation}
and
\begin{equation}
w_\mathrm{effective} := {\bar p\over\rho} =  
{\sum_{i=1}^N K_i\,w_i^2\,r^{-3w_i} \over \sum_{i=1}^N K_i\,w_i \,r^{-3w_i} }. 
\end{equation}
Note that $w_\mathrm{effective}$ can now be viewed as a position-dependent weighted average of all the $w_i$.
Finally
\begin{equation}
\Delta  = -{3\over2} 
\left(1+   
{\sum_{i=1}^N K_i\,w_i\,r^{-3w_i} \over \sum_{i=1}^N K_i\,w_i^2 \,r^{-3w_i} }
\right).
\end{equation}
So while one can still easily do various straightforward explicit calculations in this $N$-component generalized Kiselev model, one has lost many of the more compelling features of the simple one-component model.

%---------------------------------------------------------------------------------------------------------------------------------------------
\section{Rastallization}
%---------------------------------------------------------------------------------------------------------------------------------------------

Rastall gravity was introduced in 1972, some 47 years ago~\cite{Rastall:1972}. Unfortunately modern implementations of Rastall's original idea have evolved into what is merely a physically empty redefinition of parameters~\cite{Visser:2017}. These issues become particularly acute when one attempts to Rastallize the Kiselev black hole~\cite{Heydarzade:2017}. 
Effectively, the central idea of Rastall gravity is to split the ordinary conserved stress energy tensor (satisfying the ordinary Einstein equations) into two individually non-conserved pieces: 
\begin{equation}
[T_\mathrm{conserved}]^{ab} = [T_\mathrm{Rastall}]^{ab} 
+ {1\over4} \,{\lambda\over1-\lambda} \,[T_\mathrm{Rastall}] \, g^{ab}.
\end{equation}
Equivalently
\begin{equation}
[T_\mathrm{Rastall}]^{ab} = [T_\mathrm{conserved}]^{ab} - {1\over4} \,\lambda \, [T_\mathrm{conserved}] \, g^{ab}.
\end{equation}
As long as the Rastall parameter $\lambda$ satisfies $\lambda\neq1$ this can always be done, but it is merely a redefinition of what one chooses to call the stress-energy~\cite{Visser:2017}. If we now calculate the Rastall stress-energy for the one-component Kiselev spacetime in terms of the usual stress-energy we first note that
\begin{equation}
T=-\rho+3\bar p = -\rho(1-3w).
\end{equation}
Using this we obtain
\begin{equation}
\rho_{\scriptscriptstyle\mathrm{Rastall}} = \rho - {1\over4}\, \lambda\, \rho\, (1-3w) = \rho\left(1-{\lambda(1-3w)\over4}\right);
\end{equation}
\begin{equation}
(p_r)_{\scriptscriptstyle\mathrm{Rastall}} = p_r + {1\over4} \,\lambda\, \rho\,(1-3w);
\end{equation}
\begin{equation}
(p_t)_{\scriptscriptstyle\mathrm{Rastall}} = p_t + {1\over4}\, \lambda\, \rho\,(1-3w). 
\end{equation}
Consequently the absolute pressure anisotropy is invariant
\begin{equation}
(p_r)_{\scriptscriptstyle\mathrm{Rastall}} - (p_t)_{\scriptscriptstyle\mathrm{Rastall}} = p_r -p_t,
\end{equation}
while for the average pressure there is a simple shift
\begin{equation}
(\bar p)_{\scriptscriptstyle\mathrm{Rastall}} = \bar p + {1\over4} \lambda \rho(1-3w) 
= \rho\left(w+{\lambda(1-3w)\over4}\right).
\end{equation}
Furthermore
\begin{equation}
w_{\scriptscriptstyle\mathrm{Rastall}} 
= {(\bar p)_{\scriptscriptstyle\mathrm{Rastall}}\over \rho_{\scriptscriptstyle\mathrm{Rastall}} }
= {w+{\lambda(1-3w)\over4}\over1-{\lambda(1-3w)\over4} }.
\end{equation}
Finally
\begin{equation}
\Delta_{\scriptscriptstyle\mathrm{Rastall}} = 
{(p_r)_{\scriptscriptstyle\mathrm{Rastall}} - (p_t)_{\scriptscriptstyle\mathrm{Rastall}} \over (\bar p)_{\scriptscriptstyle\mathrm{Rastall}}} 
={ p_r-p_t\over (\bar p)_{\scriptscriptstyle\mathrm{Rastall}}} 
= \Delta \times {\;\; \bar p \over (\bar p)_{\scriptscriptstyle\mathrm{Rastall}} }
=
\Delta\times { w \over w+{\lambda(1-3w)\over4}}.
\end{equation}
It is easy to check that the limit $\lambda\to0$ where the Rastall parameter is set to zero is well-behaved. 
Note that the Kiselev spacetime, being anisotropic (not a perfect fluid) before Rastallization, will remain anisotropic (not a perfect fluid) after Rastallization, (As an aside, note that in reference~\cite{Visser:2017} I had performed a similar calculation for perfect fluid spacetimes; the calculation above now applies to any static spherically symmetric spacetime, including the Kiselev spacetime.)

The key physics point here is that while these formulae might superficially look somewhat impressive, they amount merely to a redefinition of parameters --- a choice as to how to split up the conserved stress-energy into two individually non-conserved pieces. 
If one starts with any spacetime satisfying the usual Einstein equations, then Rastall\-ization does not change the geometry, it is merely a book-keeping exercise applied to the stress-energy tensor. 

Specifically, since the Rastall stress-energy tensor and the usual stress-energy tensor differ only by a term proportional to the metric, the Rastallization process cannot ever affect the Hawking--Ellis classification (types I--II--III--IV) of the stress-energy tensor. (See for instance~\cite{Hawking-Ellis, Prado:2019, Prado:2018b, Prado:2018a}.) In the current context, for the spherically symmetric static Kiselev spacetime the type I stress-energy tensor remains type I. Similarly the Rainich conditions~\cite{Rainich:1925,Plebanski:1964}, and related Rainich classification of stress-energy tensors~\cite{Senovilla:2000,Bergqvist:2001,Prado:2017}, are only trivially modified by an overall shift in the Lorentz-invariant  eigenvalues, leaving the eigenvectors invariant. 

Further afield, the null energy condition (NEC) is never affected by Rastallization. However the weak, strong, dominant, flux, and trace energy conditions (WEC, SEC, DEC, FEC, TEC) are modified by a constant book-keeping offset, proportional to the trace of the stress-energy tensor. (For a general discussion see references~\cite{Hawking-Ellis,twilight,Lobo:2004,Prado:2013b,Prado:2013a,Prado:LNP}.)
Similarly the null Raychaudhuri equation and its generalizations are never affected by Rastallization, though the timelike Raychaudhuri equation and its generalizations pick up a book-keeping offset proportional to the trace of the stress-energy~\cite{Raychaudhuri:1953,Raychaudhuri:2007,Ehlers:2007,Raychaudhuri:2011}. No physics is modified by Rastallization, merely book-keeping.

\clearpage
%------------------------------------------------
\section{Discussion and Conclusions}\label{S:Conclusions}
%------------------------------------------------

Terminology is important --- only when there is widespread agreement in terminology can useful scientific progress be made. 
Having some 200 articles (over 150 of them published) use such basic concepts as ``perfect fluid'' and ``quintessence'' in a manner that is at best completely orthogonal to the usage in the bulk of the scientific community is somewhat alarming. 
While the Kiselev spacetime is an interesting toy model that does have some attractive physical and mathematical properties,  the presentation is quite often seriously deficient. 
Specifically:
\begin{itemize}
\item Do not refer to the Kiselev spacetime as perfect fluid; it isn't.
\item Do not refer to the matter in the Kiselev spacetime as quintessence; it isn't.
\item Do not try to read more into Rastall gravity than a redefinition of parameters.
\end{itemize}
I reiterate:  The fastest way to see something is wrong with the \emph{terminology} typically used to describe the Kiselev spacetime (without having to do a calculation) is simply to consider the special case $w=1/3$ with $K=-Q^2$, (where it reduces to  Reissner--Nordstr\"om spacetime), and then to note that the electromagnetic field is not a perfect fluid, nor can the electromagnetic field meaningfully be described as quintessence.

%------------------------------------------------
\acknowledgments{
This work was supported by the Marsden Fund, via a grant administered by the Royal Society of New Zealand.
}

%========================================================

%========================================================
\end{document}